\documentclass[12pt]{article}
\usepackage{graphicx}
\usepackage{amssymb,amsmath,amsfonts,multirow,rotate,color}

\usepackage{booktabs} 

\usepackage{scicite}
\usepackage{times}

\usepackage{soul}

\newcommand{\ra}[1]{\renewcommand{\arraystretch}{#1}}

\usepackage{booktabs}

\newenvironment{sciabstract}{%
\begin{quote} \bf}
{\end{quote}}

\begin{document}
\title{Information measures and cognitive limits in multilayer navigation}

\date{}

\author
{Riccardo Gallotti$^{1}$, Mason A. Porter $^{2,3}$, and Marc Barthelemy$^{1,4\ast}$\\
\\
\normalsize{$^{1}$Institut de Physique Th\'{e}orique, CEA-Saclay, F-91191, Gif-sur-Yvette, France}\\
\normalsize{$^{2}$Oxford Centre for Industrial and Applied Mathematics,}\\
\normalsize{Mathematical Institute, University of Oxford, Oxford OX2 6GG, UK}\\
\normalsize{$^{3}$CABDyN Complexity Centre, University of Oxford, Oxford OX1 1HP, UK}\\
\normalsize{$^{4}$Centre d'Analyse et de Math\'ematique Sociales, EHESS,}\\
\normalsize{190-198, Avenue de France, 75244 Paris Cedex 13, France}\\
\\
\normalsize{$^\ast$To whom correspondence should be addressed; E-mail:  marc.barthelemy@cea.fr.}
}

\maketitle

\begin{sciabstract} 

Cities and their transportation systems become increasingly complex and multimodal as they grow, and it is natural to wonder if it is possible to quantitatively characterize our difficulty to navigate in them and whether such navigation exceeds our cognitive limits. A transition between different searching strategies for navigating in metropolitan maps has been observed for large, complex metropolitan networks. This evidence suggests the existence of another limit associated to the cognitive overload and caused by large amounts of information to process.  In this light, we analyzed the world's 15 largest metropolitan networks and estimated the information limit for determining a trip in a transportation system to be on the order of 8 bits. 
Similar to the ``Dunbar number,'' which represents a limit to the size of an individual's friendship circle, our cognitive limit suggests that maps should not consist of more than about $250$ connections points to be easily readable.
We also show that including connections with other transportation modes dramatically increases the information needed to navigate in multilayer transportation networks: in large cities such as New York, Paris, and Tokyo, more than $80\%$ of trips are above the 8-bit limit.  Multimodal transportation systems in large cities have thus already exceeded human cognitive limits and consequently the traditional view of navigation in cities has to be revised substantially.

\end{sciabstract}

\section*{Introduction}

The number of ``megacities''---urban areas whose human population is larger than 10 million---has tripled since 1990~\cite{UNreport}. New York City, one of the first megacities, reached that level in the 1950s, and the world now has almost 30 megacities, which together include roughly half a billion inhabitants. The growth of such large urban areas usually also includes the development of transportation infrastructure and an increase in the number and the use of different transportation modes~\cite{Gallotti:2014}. For example, about $80\%$ of cities with populations larger than 5 millions have a subway system \cite{Roth:2012}. This leads to a natural question: Is navigating transportation systems in very large cities too difficult for humans~\cite{Roberts:2013}? Additionally, how does one quantitatively characterize this difficulty? 

It has long been recognized that humans have intrinsic cognitive limits for processing information~\cite{Miller:1956}. In particular, it has been suggested that an individual can maintain only on the order of 150 stable relationships~\cite{Dunbar:1992,Saramaki:2014}. This ``Dunbar number'' was first proposed in the 1990s by British anthropologist Robin Dunbar by extrapolating results about the correlation of brain sizes and the typical sizes of social groups for various primates. Although it is still controversial, it has been supported by subsequent studies on, e.g., traditional human societies~\cite{Dunbar:1993} and microblogging~\cite{Goncalves:2011}. 

\begin{figure*}[ht!]
\centerline{
\includegraphics[width=0.43\linewidth]{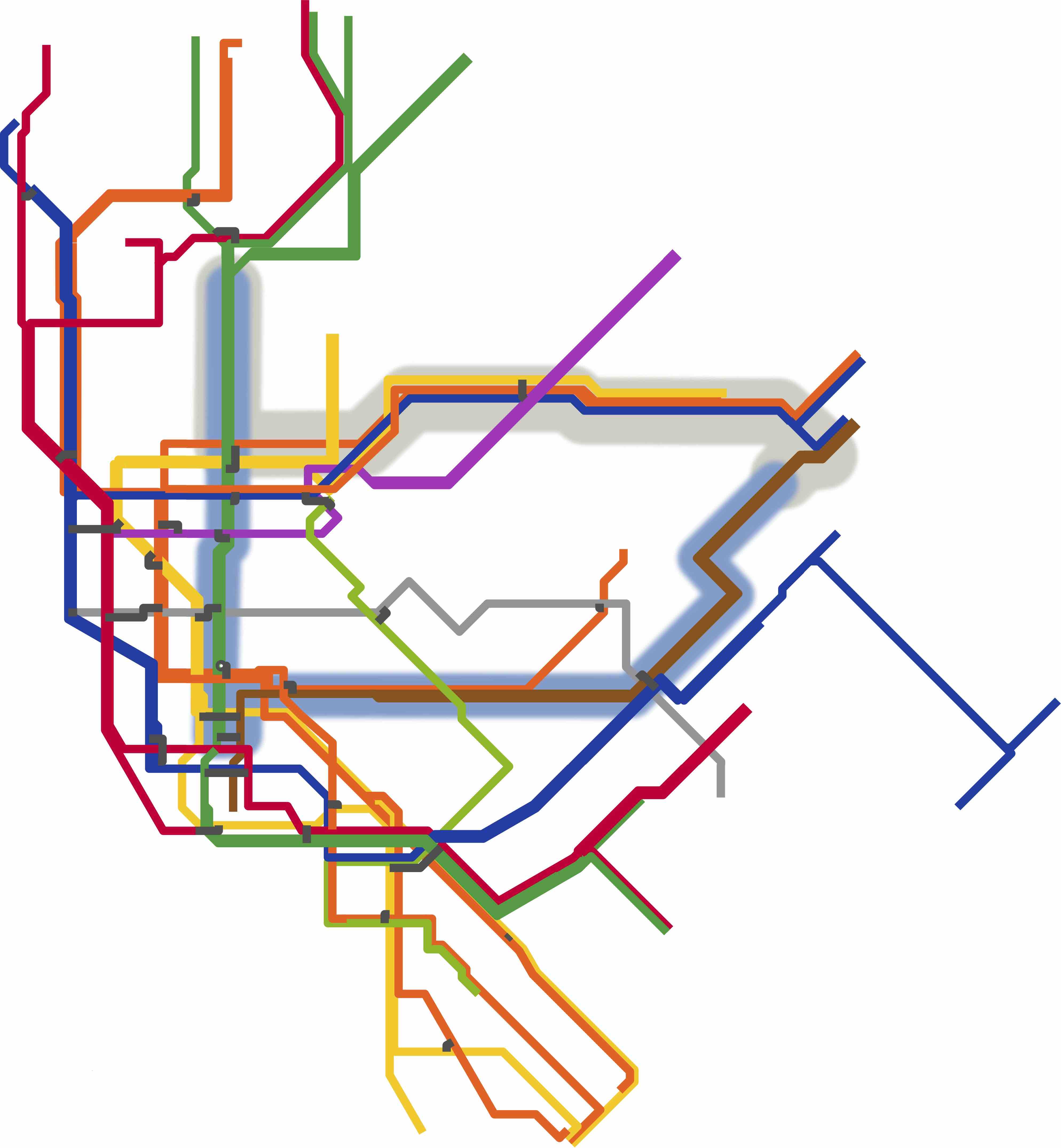} 
\qquad
\includegraphics[width=0.4\linewidth]{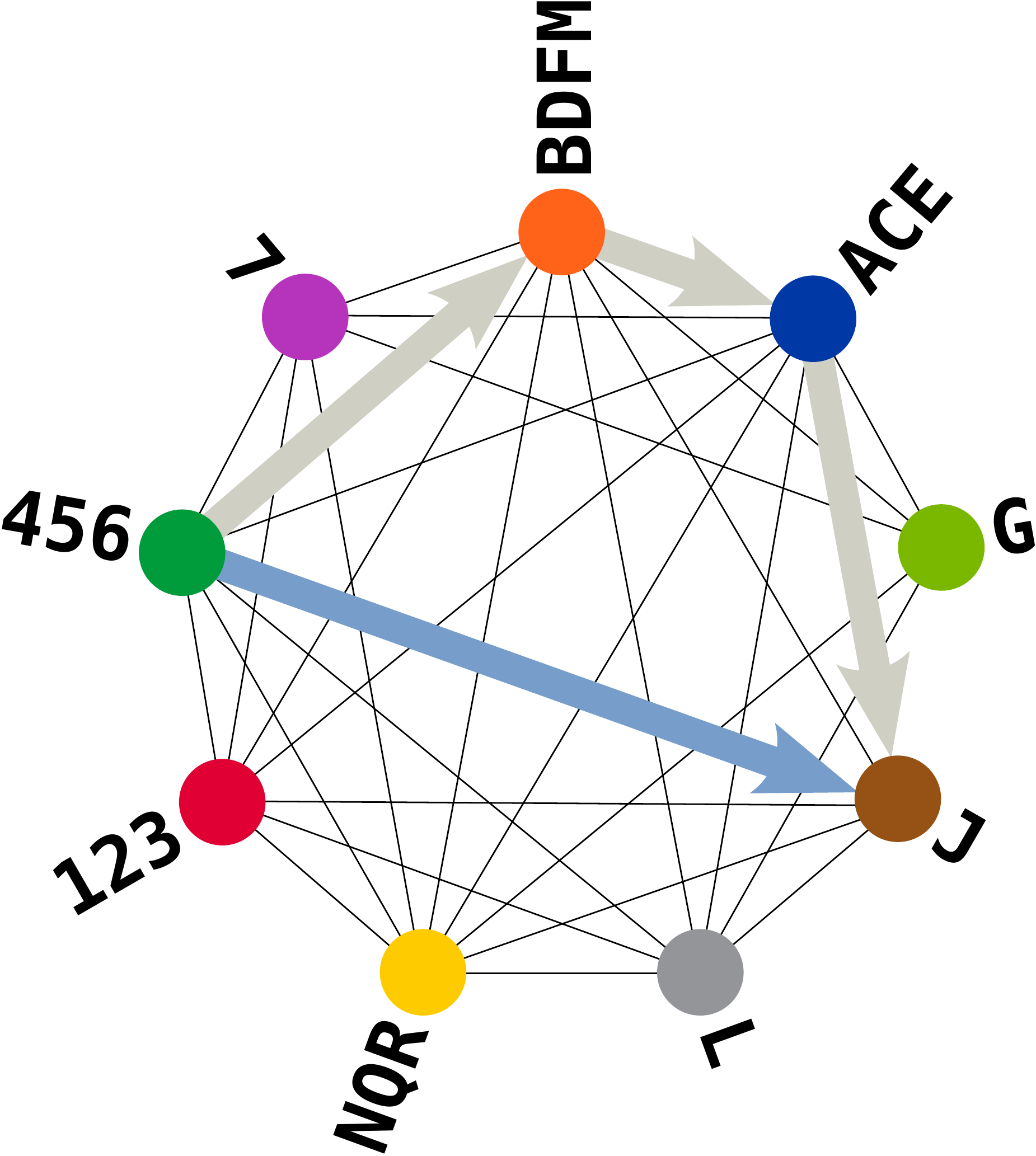} 
}
\caption{
{\bf Fastest and Simplest paths in Primal and Dual networks.} 
({\bf Left}) In the primal network of the New York City metropolitan system, the simplest path (highlighted in light blue) from 125th St. on line 5 (dark green) to 121st St. on line J (brown) differs significantly from the fastest path (highlighted in grey). There is only one connection for the simplest path ({\em Brooklyn Bridge -- City Hall / Chambers Street}) in Lower Manhattan. In contrast, the fastest path needs three connections (5$\to$F$\to$E$\to$J). We compute the fastest path using travel times from the Metropolitan Transportation Authority (MTA) Data Feeds (see Materials and Methods). We neglect walking and waiting times.
({\bf Right}) In the dual space, nodes represent routes and edges represent connections. A simplest path in the primal space is defined as a shortest path with the minimal number of edges in the dual space (light blue arrow). It has a length of $C=1$ and occurs along the direct connection between line 5 (dark green node) and line J (brown node). The fastest path in the primal space has a length of $C=3$ (grey arrows) in the dual space, as one has to change lines three times. (We extracted the NYC metro schematic from a map that is publicly available on Wikimedia Commons~\cite{nycMapCommons}.)  
}
\label{fig1}
\end{figure*}

When navigating for the first time between two unfamiliar places and having a transportation map as one's only support, a traveler has to compare different path options to find an optimal route. Differently from the case of the schematization of partially familiar routes~\cite{Srinivas:2007}, here the traveler does not need to simultaneously visualize the whole route; it is sufficient to identify and keep track of the position of the connecting stations on the map. Therefore, a first important point to consider is that humans can track information on a maximum of about four objects in their visual working memory \cite{Luck:1997}. This implies that a person can easily keep in mind the key locations (origin, destination, and connection points) for trips with no more than two connections (which corresponds exactly to four different points). In addition, recent studies on visual search strategies~\cite{Burch:2014b,Burch:2014a} show a transition in search strategies between the simple cases of the Stuttgart and Hong Kong metropolitan networks and the case of Paris, which has one of the most complicated transportation networks in the world. The time needed to find a route in a transportation network grows with the complexity of its map, and the pattern of eye fixations also changes from following metro lines to a random scattering of eye focus all over the map~\cite{Burch:2014b}. A similar transition from directional to isotropic random search has been observed for visual search of hidden objects when increasing the number of distractors~\cite{Credidio:2012}. The ability to manage complex ``mental maps'' is thus limited, and only extensive training on spatial navigation can push this limit with morphological changes in the Hippocampus~\cite{Maguire:2000}. Human-constructed environments have far exceeded these limits, and it is interesting to ask whether there is a navigation analog of the Dunbar number and a cognitive limit to human navigation ability, such that it becomes necessary to rely on artificial systems to navigate in transportation systems in large cities. If such a cognitive limit exists, what is it? In this paper, we answer this question using an information perspective~\cite{Miller:1956} to characterize the difficulty to navigate in urban transportation networks. We use a measure of ``information search'' associated to a trip that goes from one route to another~\cite{Rosvall:2005}. In most networks, many different paths connect a pair of nodes, and one generally seeks the fastest path that minimizes the total time to reach a destination. However, it tends to be more natural for most individuals to instead consider a ``simplest'' path, which has the minimum number of connections~\cite{Viana:2013} (see Fig.~\ref{fig1}).

% FIGURE 1

Rosvall {\it et al.}~\cite{Rosvall:2005} proposed a measure for the information that is needed to encode a shortest path from a route $s$ to another route $t$. However, the amount of necessary information can depend strongly on the initial and final nodes, and we consider a trip from an origin node $i$ in route $s$ to a destination node $j$ in route $t$. This trip is embedded in real space and among all possible simplest paths~\cite{Rosvall:2005} (which need not to be unique), we pick the fastest one $p(i,s;j,t)$, which can differ from an actual fastest path between $i$ and $j$ (see the left panel of Fig.~\ref{fig1}). 
For computing the travel time of a trip, we neglect the contribution of walking and waiting times~\cite{Gallotti:2014}. However, the choice of a simplest path already tends towards minimization of such transfer costs, which strongly influence a traveler's decisions~\cite{Wardman:2004}.

The total information for knowing the fastest simplest path is 
\begin{equation}
\label{definitionS}
	S(i,s;j,t) =  -\log_2\left (\frac{1}{k_s}\prod_{n\in p(i,s;j,t)}\frac{1}{k_n-1}\right)\,,
                      %=\log_2{(k_s)} +  \sum_{n\in p(i,s;j,t)} \log_2{(k_n-1)}\,,
\end{equation}
where $p(i,s;j,t)$ is the sequence of routes needed for connecting $i$ in route $s$ to $j$ in route $t$. The term $k_s$ is the number of routes connected to $s$, and along the path, we have the choice between $k_n-1$ routes. The idea behind Eq.~(\ref{definitionS}) is that when tracking a trip along a map (with the
eyes or a finger), the connections that one has to exclude represent---similarly to the number of distractors in visual search tasks~\cite{Credidio:2012}---the information that has to be processed and thus temporarily stored into working memory~\cite{Baddeley:2003}. One can therefore construct the measure of entropy (\ref{definitionS}) as a proxy for the accumulated cognitive load that is associated to the trip, and it is analogous to the total amount of load experienced during a task~\cite{Paas:2003}. For this reason, this measure of entropy seems to be appropriate for estimating the information limit associated to the observed transition in the visual search strategy~\cite{Credidio:2012,Burch:2014b}.

From a map user's perspective, the existence of several alternative simplest paths is not necessarily a significant factor, as one only needs a single simplest path for successful transportation from origin to destination. Consequently, we use the entropy in Eq.~(\ref{definitionS}) rather than the one proposed in Ref.~\cite{Rosvall:2005}. (See the Materials and Methods for further discussion.) To produce a single summary statistic for a path, we average $S(i,s;j,t)$ over all nodes $i\in s$ and $j\in t$ (we denote this mean using brackets $\langle\cdot\rangle$) to obtain
\begin{equation}
	\bar S(s,t) = \langle S(i,s;j,t)  \rangle\,,
\end{equation}
which is the main quantity that we use to describe the complexity of a trip (see Fig.~5) and which will allow us to extract an empirical upper limit to the information that a human is able to process for navigating. 

\section*{Results}

We use the measure $\bar S(s,t)$ to characterize the complexity of the 15 largest urban metropolitan systems in the world. From these results, we then extract an empirical upper limit of about $8$ bits to the information that a human is able to process for navigating. We then apply this threshold in calculations for multimodal transportation networks and demonstrate that most trips in large cities exceed human cognitive limits.

\subsection*{Information Threshold}

% FIGURE 2
\begin{figure*}[ht!]
\centerline{
\includegraphics[width=0.5\linewidth]{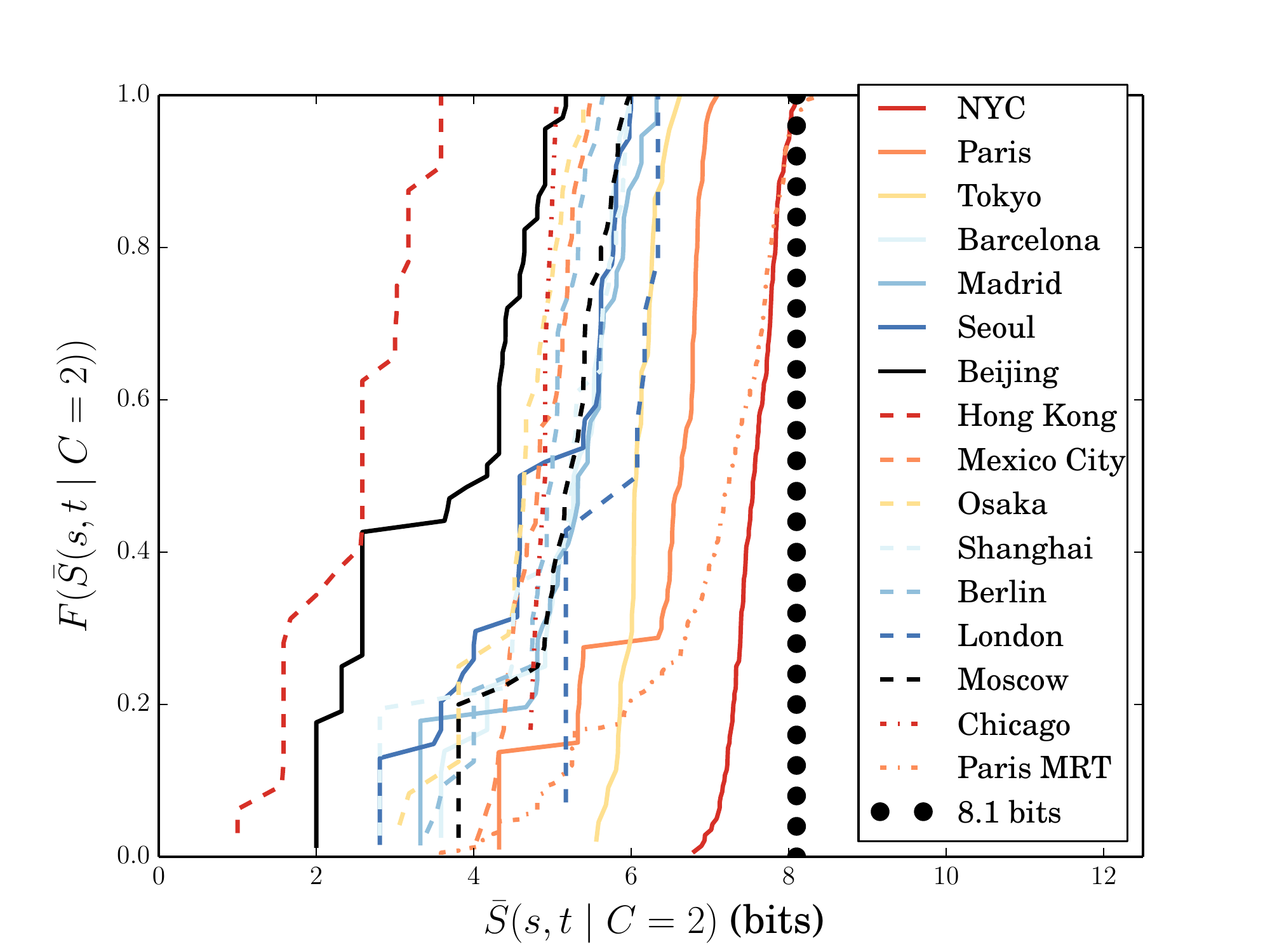} 
\quad
\includegraphics[width=0.5\linewidth]{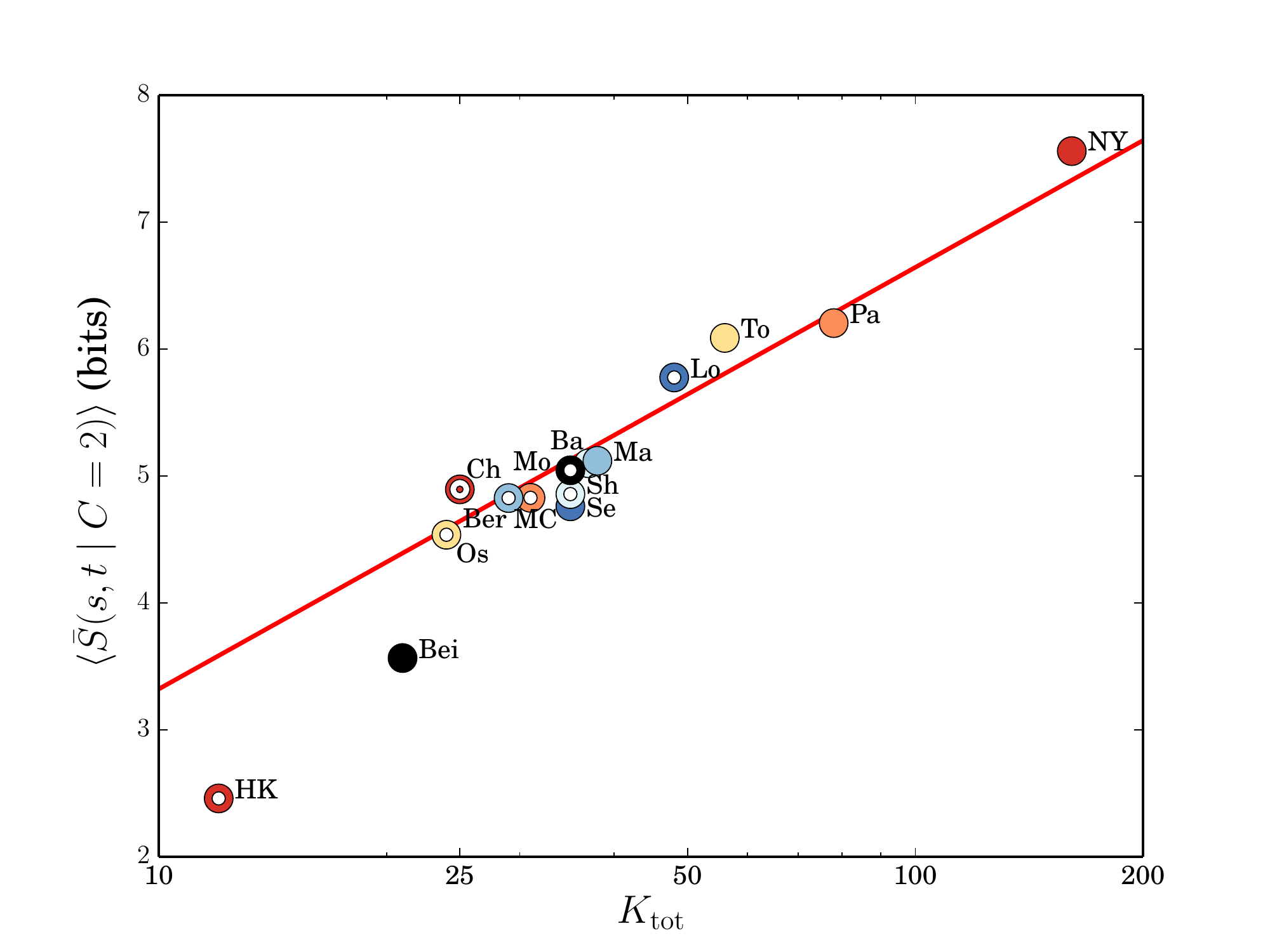} 
}
\caption{
{\bf (Left) Information threshold.} Cumulative distribution of the information needed to encode trips with two connections in the 15 largest metro networks. The largest value occurs for the New York City metro system (solid red curve), which has trips with a maximum of $S_{\mathrm{max}} \approx 8.1$ bits. Among the 15 networks, the Hong Kong (dashed, red) and Beijing (solid, black) metro networks have the smallest number of total connections and need the smallest amount of information for navigation. The Paris MRT (Metro, Light Rail, and Tramway) network (orange, dash-dotted) from the official metro map~\cite{planMetro}. includes three transportation modes (which are managed by two different companies) and reaches values that are similar to those in the larger NYC Metro.
{\bf (Right) Information threshold versus total number of connections in the dual space.} This plot illustrates that the mean amount of information that is needed to encode trips with two connections is strongly correlated with the total number of connections in the dual network, as can be predicted for a square lattice (see Materials and Methods). See Table~1 for the definitions of the abbreviations. The color code is the same as in the left panel, and the red solid line represents the square-lattice result $\bar S(s,t|C=2) = \log_2\left( K_{\mathrm{tot}}\right)$. This relationship permits one to associate the information threshold $S_{\mathrm{max}}$ with the cognitive threshold $T= 2^{S_{\mathrm{max}}}$, which one can interpret as a maximum number $T$ of intra-route connections that can be represented on a map.
}
\label{fig2}
\end{figure*}

The values of $\bar S(s,t)$ in a network tend to grow with the number $C$ of connections that appear in a simplest path as well as with the meanmean degree $\langle k \rangle$ of the nodes in the dual space (see Figs.~6 and 7). Note that the latter is related to the total number of connections in a network. Adding new routes can thus have a negative impact from the information perspective. Although new routes can be useful for shortening the simplest paths for some $(s,t)$ pairs, new connections simultaneously increase the mean degree of a network and can make it more difficult to navigate in a network. 
We thus want to estimate the maximum possible information that an individual can reasonably process to navigate in a transportation system. For that purpose, we consider the world's 15 metro networks with the largest number of stations. The characteristics of metro networks were examined in previous papers~\cite{Latora:2001,Angeloudis:2006,Lee:2008,Derrible:2010,Roth:2012,Rombach:2014,Louf:2014}, and navigation strategies have been considered in transportation networks~\cite{shl-prl,Hoffmann:2013,DeDomenico:2014}. For each network, we consider the shortest simplest paths with $C=2$ connections. This corresponds to paths that use $3$ different lines: such a path starts from a source route $s$, connects to an intermediate route $r$, and then connects to a destination route $t$. There are two distinct reasons for this choice: (i) the limit of four objects in the visual working memory~\cite{Luck:1997}; (ii) in most of the 15 cities, two connections correspond to the diameter of the dual network. From a map user's perspective, after having checked that each pair of consecutive stations are connected by a direct line, the locations to keep in mind are the origin, the destination, and connecting stations. These nodes correspond to the places that one has to  ``highlight'' on the map to record the trajectory. The capacity of visual working memory thus allows one to easily keep in mind only trajectories with 2 connections, leading to a total of 4 stations. Interestingly, the value 2 is also the diameter of the dual network in most of the metropolitan systems. In such situations, all pair of nodes can be connected by paths with at most 2 connections, staying below the working memory limit of humans.

In the left panel of Fig.~\ref{fig2}, we show the cumulative distribution of entropies $\bar S(s,t|C=2)$ for these 2-connections paths. We find that the New York City metro system is the largest and most complex metropolitan system in the world; it has a maximal value of $S_{\mathrm{max}} \approx 8.1\approx \log_2(274)$ bits. Paris' transportation system reaches a similar value if one takes into account the light rail and tram system in the multilayer Metro-Rail-Tramway (MRT) network displayed in the official metro map~\cite{planMetro}.

Navigation in such large networks is already nontrivial~\cite{Burch:2014a}, and it has been observed that there is an eye-movement behavioral transition when the system becomes too large (i.e., when there are too many connections)~\cite{Burch:2014b}. The value $S_{\mathrm{max}}$ for trips with two connections thus provides a natural limit, above which human cognitive capabilities are challenged and for which it becomes extremely difficult to find a simplest path. We thus make the reasonable choice to take $S_{\mathrm{max}}$ as the cognitive limit for public transportation: a human needs an information entropy of $\bar S(s,t) \leq S_{\mathrm{max}}$ to be able to navigate in a network successfully without assistance from information technology tools.

\begin{figure*}[ht!]
\centerline{
\includegraphics[width=0.8\linewidth]{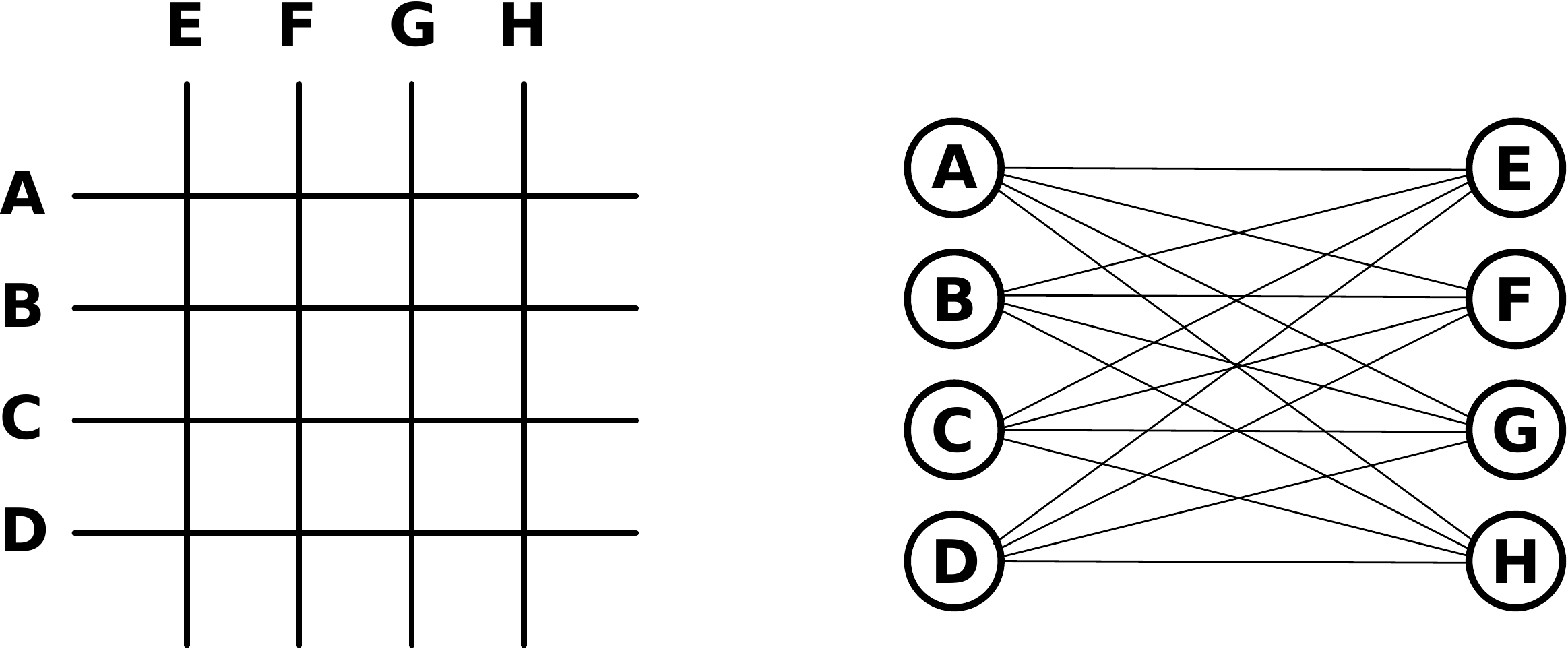} 
}
\caption{{\bf (Left) Primal and (right) dual networks for a square lattice.} In this example, the lattice has $N=8$ routes. Each route has $k=N/2=4$ connections, so the total number of connections is $K_{\mathrm{tot}} = k^2 = 16$. In the dual network, the four East-West routes (A,B,C,D) and the four North-South routes (E,F,G,H) yield a graph with a diameter of 2.}
\label{lattice}
\end{figure*}

To gain a physical understanding for the cognitive limit $S_{\mathrm{max}}$, we estimate $S(s,t|C=2)$ for a regular 
lattice (like the one in Fig.~\ref{lattice}) with $N$ lines that are each connected with $N/2$ other lines (i.e., $k_r = N/2$ for all $r$). This choice of a lattice is justified by the results in Ref.~\cite{Roth:2012} that most large metropolitan transportation networks consist of a core set of nodes with branches that radiate from it. The core is rather dense and has a peaked degree distribution, so it is reasonable to use a regular lattice for comparison. In the dual space of the regular lattice, the degree $k_s$ of route $s$ is equal to $N/2$, and we thus obtain
\begin{align}
\nonumber
	\bar S(s,t|C=2) &= \log_2[k_s (k_r-1)] \\
&\approx \log_2(\langle k\rangle^2) = \log_2\left(\sum_{i=1}^N k_i/2\right)\,,
\label{eqS2}
\end{align}
where $\langle k\rangle$ denotes the mean degree. The last equality in Eq.~(\ref{eqS2}) comes from the relation for the total degree of a regular lattice: $\sum_{i=1}^N k_i = \langle k\rangle N = 2\langle k\rangle^2$. The key quantity for understanding $S_{\mathrm{max}}$ is therefore the total number of undirected connections $K_{\mathrm{tot}} = \sum_{i=1}^N k_i/2$ in the dual space (also see Table~1). As we indicated in Eq.~(\ref{eqS2}), this is identical to the square of the mean degree $\langle k\rangle^2$ in a lattice. For Paris, for example, we obtain $\langle k\rangle \approx 9.75$, which leads to $9.75^2\approx95$ connections for the corresponding lattice. The actual Paris metropolitan network has a total of 78 connections, and the difference comes from the fact that the real network is not a perfectly regular lattice.  At this stage, it is important to make two remarks. First, the apparently paradoxical fact that the total information grows with size for regular lattices while intuitively the complexity for finding a path stays constant is specific to the case in which there is an ``algorithm'' to find the route. Indeed, in perfectly regular rectangular (i.e., Cartesian) grids, one needs to make at most two turns to find a desired route. Second, drawing a parallel with an individual navigating a bifurcating tree from the root to a terminal node who is traversing 8 bifurcation points (and thus 255 internal nodes), the value of 8 bits as a limit appears to be consistent with Miller's ``magic number''~\cite{Miller:1956,Baddeley:1994}. ``Miller's law'' limits also the number of binary decisions that can be memorized in sequence to a value that has been observed to be in the range $7\pm2$.

% FIGURE 3

We test Eq.~(\ref{eqS2}) in the right panel of Fig.~\ref{fig2} for the 15 largest metropolitan networks. Our calculation shows that the total degree in the dual space, which is related to the total number of connections in a network, is the main ingredient for understanding the information entropy of these systems. We also test this relation for the temporal evolution of the Paris metro network and we show that the number of connections $\sum_{i=1}^N k_i/2$ scales as $(N/2)^2$ for the historical growth from $N=1$ to $N=14$ routes (see Fig.~8). Equation~(\ref{eqS2}) allows us to translate the information limit of $8$ bits to a limit on the number $T$ of intra-route connections ($S =\log_2(T))$. The value of $T$  also corresponds to the number of distractors to be excluded for the most complex trips (with $C(s,t)=2$). This process of exclusion thus demands progressive information integration, which causes a cognitive overload.

Evidence for the existence of such a cognitive threshold is the change in search strategy observed in eye-tracking experiments~\cite{Credidio:2012,Burch:2014b}. The value $T \approx 250$ represents the worst-case scenario in the world's largest metropolitan network. It thus overestimates the values at which the transition occurs.  Indeed, the Paris MRT network, for which the strategy change was observed in \cite{Burch:2014b}, has $K_\mathrm{tot} \approx 162$. 
. 
It is interesting that $T$ has a similar order of magnitude as the Dunbar number, an extensively studied cognitive limit for the size of a friendship's circle, and which seems to lie in a range between $100$ and $200$. (See, for example, \cite{Goncalves:2011} for a recent discussion of this topic.).

%\clearpage

\subsection*{Effect of Multimodal Couplings on the Information}

%FIGURE 4
\begin{figure*}[h!]
\centerline{
\includegraphics[width=0.9\linewidth]{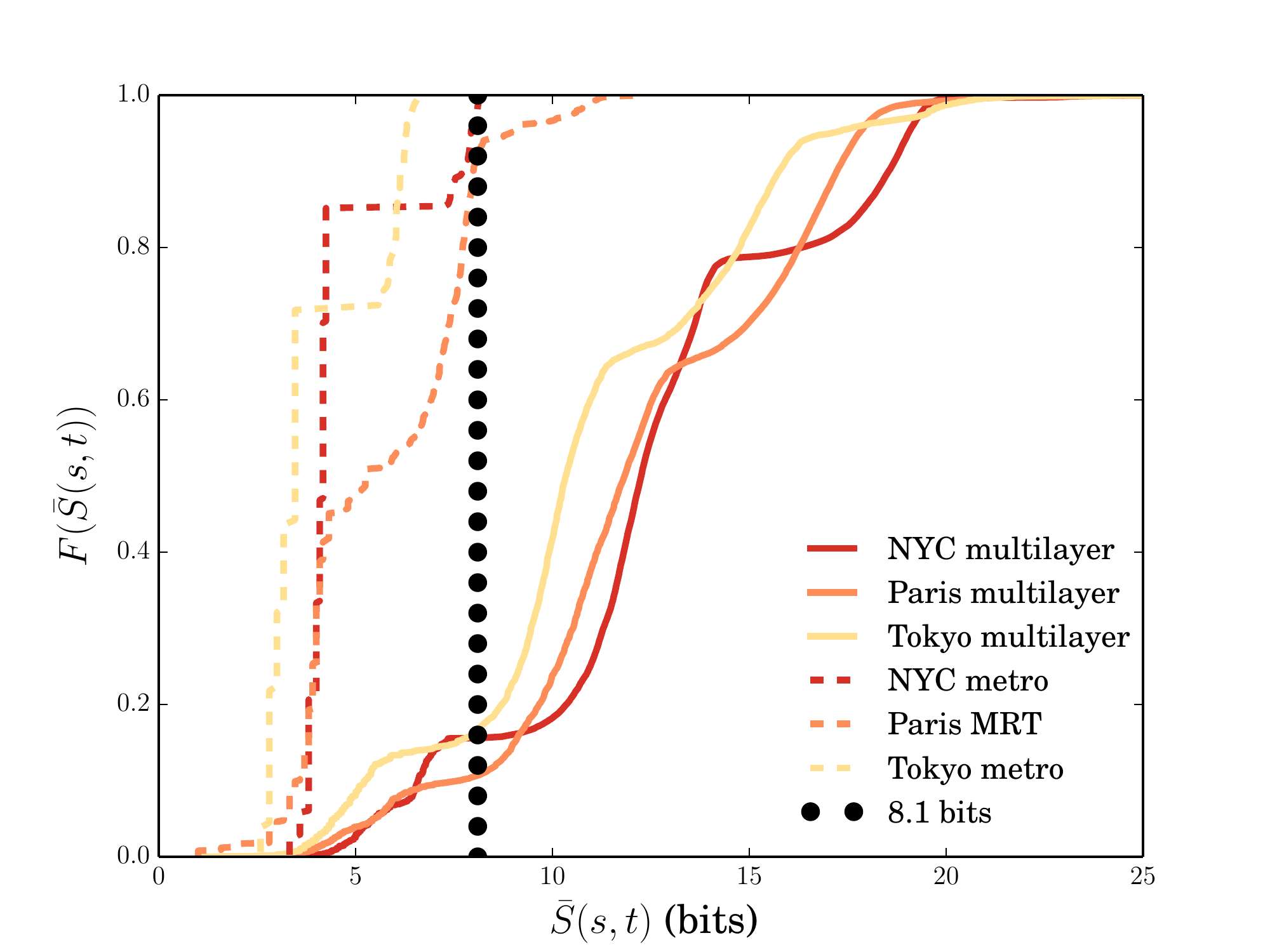} 
}
\caption{
{\bf Information entropy of multilayer networks}. The solid curves represent the cumulative distributions of $\bar S(s,t)$ for multilayer networks that include a metro layer for New York City, Paris, and Tokyo. (We associate one layer to bus routes and another to metro lines.) Most of the trips require more information than the cognitive limit $S_{\mathrm{max}}\approx8.1$. The fraction of trips under this threshold are $15.6\%$ for NYC, $10.7\%$ for Paris, and $16.6\%$ for Tokyo. The dashed curves are associated to all possible paths in a metro layer; in this case, the amount of information is always under the threshold, except for Paris, which includes trips with $C=3$. (See Table~3). Note that the threshold value lies in a relatively stable part of all three cumulative distributions, which suggests that our results are robust with respect to small variations of the threshold.
}
\label{fig4}
\end{figure*}

In our discussion above, we estimated the cognitive threshold for the most complex paths in the 15 largest metropolitan networks. We now consider the effect of including other transportation modes (e.g., buses, trams, etc.). The effects of inter-modal coupling are significant~\cite{Morris:2012,Gallotti:2014}, and the natural framework is a multilayer network~\cite{Kivela:2014,Boccaletti:2014}, which associates each transportation mode with a different ``layer'' in a network and where interchanges (i.e., connection points) between different modes are represented by inter-layer edges. As we discuss in the Materials and Methods section, a major difficulty comes from obtaining data for different modes for a given city, as this type of data is not currently available in general. However, we were able to combine multiple data sources for large cities on three different continents: New York City, Paris, and Tokyo (see Table~2).

The distribution of $\bar S(s,t)$ is a superposition of peaks associated to different values of
$C$ (see Fig.~6). By comparing the distribution of $\bar S(s,t)$ for the bus
monolayer network and the full multilayer transportation network, we can distinguish two
competing effects of multimodality: 
(i) it tends to reduce the number $C$ of connections and thereby reduces $\bar S$; and (ii) it increases $\bar S$, because new routes increase the node degrees in the dual space. (See Fig.~7) However, these two contributions do not compensate each other. In Fig.~\ref{fig4}, we show the cumulative distribution $F(\bar{S}(s,t))$ of information entropy values for the New York, Paris, and Tokyo multimodal (i.e., multilayer) transportation systems.

We find that less than approximately $17\%$ (the maximum value obtained for Tokyo) of the trips are below the threshold $S_{\mathrm{max}}$. These results imply that more than $80\%$ of the trajectories in the complete public transportation networks of these major cities require more information than the most complicated trajectory in the largest metro networks. As we show in the Supplementary Materials, the other $20\%$ correspond to pairs of nodes for which the trip has essentially one connection (for NYC) or at most two connections (for Paris and Tokyo), as the simplest paths that carry a small amount of information are those that avoid using too many major hubs. (See Table~4 and Fig.~9). The number of connections acting as distractor for the case of the Paris MRT is already so large that it has a crucial impact in the route search (it takes roughly $30$ seconds on average for such as search~\cite{Burch:2014a}), and the complexity of the bus layer (and therefore of the coupled metro and bus system) will therefore exceed the human capacity. Consequently, traditional maps that represent all existing bus routes have a very limited utility. This result thus calls for the need of thinking about a user-friendly way to present and to use bus routes. For example, unwiring some bus-bus connections lowers the information, and leads to the idea that a design centered around the metro layer could be efficient. Further work is however needed to reach an efficient, 'optimal' design from a user perspective. 

Finally, we note that one can also think of finding an optimal path by separately considering different transportation modes. The price to pay for such a strategy would be to settle for suboptimal trajectories (which are neither simplest nor shortest), as an optimal trajectory is, by definition, computed on a full multilayer transportation network.

\section*{Discussion}

Human cognitive capacity is limited, and cities and their transportation network have grown to the point that they have reached a level of complexity that is beyond human processing capability to navigate in them. In particular, the search for a simplest path becomes inefficient when multimodality is important and when a transportation system has too many interconnections. This occurs because of interconnections, which play two crucial roles in the search for a path: they are both targets and distractors. The identification of possible interchange points is a key and extremely time-consuming passage in the route-finding task~\cite{Burch:2014b}. As with the case of hidden objects~\cite{Credidio:2012}, one can represent the difficulty of the search using the number of distractors, which for a map are all possible interchanges. We have found in the largest cities that the addition of bus routes with maps that are already too complicated to be easily used by travelers implies that the cognitive limit to urban navigation is exceeded for multimodal transportation systems. 

We have estimated the cognitive threshold value to be of $T\approx 250$ connections (i.e., approximately 8 bits), which represents the worst-case scenario for the most complex trips in the largest networks. This exceeds the behavioral transition, beyond which humans have difficulties to navigate on their own.

Remarkably, we note that $T$ has a similar interpretation and order of magnitude as the Dunbar number, and our result can thus be construed as  evidence to suggest the existence of a  ``transportation Dunbar number'' for processing complex maps. Although the value of $T$ represents an upper bound, it allows us to demonstrate the existence of the huge gap between the amount of information that a human can process easily ($K_{\mathrm{tot}}  < 250$) and the amount that is contained in practice in multimodal transportation networks in large cities ($K_{\mathrm{tot}} > 1800$, see Table~2). Indeed, the growth of transportation systems has yielded networks that are so entangled with each other and so complicated that a visual representation on a map becomes too complex and ultimately useless. In particular, our results imply that maximizing the number of intersections between lines, which indeed minimizes transfers, is contrary to the goal of having an easily usable transportation system. The information-technology tools provided by companies and transportation agencies that help people to navigate in transportation systems will soon become necessary in all large cities. Our analysis highlights the fact that humans need to integrate navigation complexity and seek new solutions that will help for navigation in megacities. Redesigning maps and representations of transportation networks~\cite{spidermaps} as well as improving information-technology tools~\cite{DeDomenico:2015} that help to decrease the amount information below the human processing threshold thus appear to be crucial for an efficient use of services provided by transportation agencies.

\section*{Materials and Methods}

\subsection*{Data for the World's 15 Largest Metropolitan Transportation Networks}

The data that we use was extracted from Wikipedia~\cite{wikipedia}. It describes the lines in the metropolitan networks as they were in 2009, and the data were used in a previous publication~\cite{Roth:2012}. The data for each metropolitan system yields a spatial network~\cite{Barthelemy:2011}; each station has geographical coordinates, and the edges connect consecutive stations on a metro line. The diagnostic that we use for determining a shortest simplest path is the sum of the traveling times between stops.

\subsection*{Paris}

The data that we use for constructing the Paris multilayer network~\cite{Kivela:2014,Boccaletti:2014} come from the R\'egie Autonome des Transports Parisiens~\cite{RATP} and the Soci\'et\'e Nationale des Chemins de fer Fran\c{c}ais~\cite{SNCF}. Both data sets are provided in GTFS format. We extract the bus layer from the RATP data. To reproduce all of the information available in the Paris Metro Map~\cite{planMetro}, we merge the metro and tramway data from the RATP with the light rail and tramway data from the SNCF. This aggregation gives what we call the ``metro-rail-tramway'' (MRT) layer, which we use as a single metro layer. We distinguish services that go in opposite directions or towards different branches, and we are able to identify them as a particular route because they share the same short name. We use the services that were available on 
Monday 26/5/2014, and we exclude any bus route that is completely subsumed by another route. 

The RATP data provide extremely detailed transfer times between stops. We use this information to reconstruct connections between routes. With the exception of connections in the central station Ch\^atelet-Les Halles, we ignore transfers that take longer than 8 minutes and consider the two corresponding nodes as different entities. By studying the cumulative distribution for the walking distances of these connections data, we find that the 99th percentile corresponds roughly to $d_w = 250 m$. More precisely, $F(250 m) \approx 0.986$. Motivated by this calculation, we allow a maximum walking distance of t $d_w = 250 m$ for the coarse-graining procedure that we need to construct the multilayer networks in New York City (NYC) and Tokyo.

For Paris, the transfer data allow a more accurate method than for NYC and Tokyo. We construct a transfer network in which (i) a transfer is defined between two stops and (ii) two stops are located in the same geographical position. We associate a single node to each connected component of the transfer network, and each isolated stop constitutes a single node. We weight the edges using travel times. 
Relatively large connection areas emerge from our choices. They correctly reflect possible choices that are suggested to travelers on the Paris metro map~\cite{planMetro} (via the representation of intra-station walking paths as white edges).

Finally, we also couple nodes if (i) they are closer than $d_w = 250$ meters and if they come from RATP data and are labeled with the same name or (ii) they come from SNCF data (where transfer times between lines were not provided). Intra-route connections are possible if routes share the same node. As for NYC, the trip length is equal to the sum of the mean travel times between consecutive stops.

\subsection*{New York City}

We construct the NYC multilayer network with data from the Metropolitan Transportation Authority (MTA) Data Feeds~\cite{MTA}, which provide a snapshot of their service for different days using the General Transit Feed Specification (GTFS)~\cite{GTFS}. We use the services that were available on Monday 1/12/2014. For the metro layer, we exclude six routes that represent shuttles or other special services.

For the bus layer, we integrate the data from the five NYC boroughs with multi-borough data of the subsidiary MTA Bus Company. We perform a coarse-graining procedure in which (i) we couple bus stops to metro locations if they lie within a walking distance of $d_w = 250$ meters and then (ii) combine bus stops into a single entity if their distance is less than $d_w$. We therefore allow intra-route connections if two routes share the same stop (where coupled stops count as a single stop). As with Paris, the trip length is given by the sum of the mean travel times between consecutive stops.

\subsection*{Tokyo}

The Tokyo metro data set, which we have extracted from Wikipedia, allows us to study another of the world's largest metropolitan networks~\cite{wikipedia}. The bus data for Tokyo (for 2010) is provided freely by the Japanese Ministry of Land, Infrastructure, Transport, and Tourism~\cite{Tokyo}. We use only Toei Bus lines~\cite{toei}, which serve central Tokyo.

This data set associates stops to bus routes, but it gives no information about the topology of the line. We reconstruct an approximate topology by generating a minimal spanning tree~\cite{Clark:1991} among all stops of a route. We define the intra-route connections using the same coarse-graining procedure as for  NYC.  We again combine nodes that are within a walking distance of $d_w=250$ meters from each other. We have no information on the travel times along the edges, so we estimate them from the trip length and using typical transportation speeds for the bus and metro in the Paris data. Because one can approximate each layers' speed by a log-normal distribution, we use the log-average $\bar v = \exp{\langle \log(v)\rangle}$ as the typical speed associated to each mode of transport. The values that we find are $v_B \approx 14.0$ km/h) for the bus and  ($v_M \approx 23.4$ km/h) for the metro.

\subsection*{Definitions of Paths and Information Entropy}

From the perspective of information processing, one can quantify the difficulty of navigating in an
urban transportation network using a measure of ``search information'' $S$~\cite{Rosvall:2005}. 
This measure represents the amount of information that is needed for encoding a path from a route $s$ to a route $t$ that follows a simplest path. We define a \emph{simplest path} as a shortest path in
the dual space of a transportation network. In the present context, the dual space (which is also
called ``P-space''~\cite{Barthelemy:2011}) is the network in which the nodes represent routes and the edges represent possible intersections (or connections between different lines) among those routes. There can be many degenerate simplest paths between two routes, and one can define the information entropy computed on these paths to be~\cite{Rosvall:2005} 
\begin{equation}\label{thisexpression}
	S^D(s,t) = -\log_2 \left[ \sum_{\{ p(s,t)\}} \left( \frac{1}{k_s} \prod_{n \in \{p(s,t)\}} \frac{1}{k_n-1} \right) \right]\,,
\end{equation}
where the set $\{ p(s,t)\}$ includes all $D(s,t)$ degenerate simplest paths between routes $s$ and $t$. The quantity $k_s$ indicates the total number of connections that emanate from route $s$. On a path, one has a choice
between $k_n-1$ routes (i.e., excluding the route whence the path emanates).

When one is seeking an optimal trajectory, the degeneracy is not necessarily a
significant factor, and we focus on a trip from a
node $i$ to another node $j$ in real space. Among all of the degenerate simplest paths, we pick a shortest
one $p(s,t,i,j)$ and consider its entropy
\begin{equation}
	S(i,s;j,t) =  \log_2{(k_s)} +  \sum_{n\in \{p(s,t,i,j)\}} \log_2{(k_n-1)}\,.
\end{equation}
Averaging over all nodes $i\in s$ and $j\in t$ yields the main
quantity that we used in the main text:
\begin{equation}
	\bar S(s,t) = \langle S(s,t,i,j)  \rangle\,.
\end{equation}
In theory, it is also possible to weight this mean using real flows. Unfortunately, for this particular case, traditional data sources are insufficient, as they tend to describe the mobility of commuters, who do not necessarily rely on maps for their daily journeys.
If we assume that all of the degenerate paths contribute equally to the entropy, then we obtain the following approximate relation between the entropies: 
\begin{equation}
	\bar S(s,t) \approx S^D(s,t) + \log_2[D(s,t)]\,.
\label{SsimshoVSrosvall}
\end{equation}
Deviations from the (approximate) equality (\ref{SsimshoVSrosvall}) indicate differences between the
various degenerate paths (see Fig.~10).

From a map user's perspective, the existence of several alternative
simplest paths is not necessarily a significant factor, as one only
needs a single simplest path for successful transportation from origin
to destination. The natural choice is the fastest simplest path
path, which is granted as unique for continuous travel-times. Consequently, we use the entropy in
Eq.~\ref{definitionS} rather than the one proposed in
Ref.~\cite{Rosvall:2005}

\section*{Acknowledgements}

We thank M.~Kivel\"a for helpful comments on an early draft of this paper. We thank P. Broderson, A. Sol\'e Ribalta and the anonymous reviewers for useful comments. RG thanks L.~Alessandretti for suggesting the use of SNCF data. MB thanks Dr. Y. Shibata for helping to find and process the Tokyo data. All authors are supported by the European Commission FET-Proactive project PLEXMATH (Grant No. 317614). The authors declare no competing financial interests.

%%%%%

%\clearpage

\section{Supplementary Materials}

\begin{table}[h!]
\centering
\ra{1.3}
\begin{tabular*}{.8\textwidth}{@{\extracolsep{\fill}}lccccccc@{}}
\toprule[1pt]
City & Abbreviation & Nodes & Edges & $N$ & $K_{\mathrm{tot}}$ & P-diameter \\
\hline
New York City	& NY 	& 433 & 497 & 22 & 161 & 2\\
Paris	 	  	& Pa		& 299 & 355 & 16 & 78   & 3\\
Tokyo	  	& To		& 217 & 262 & 13 & 56   & 2\\
London	  	& Lo		& 266 & 308 & 11 & 48   & 2\\
Madrid	  	& Ma		& 209 & 240 & 12 & 38   & 2\\
Barcelona	  	& Ba		& 139 & 165 & 11 & 37   & 2\\
Moscow	  	& Mo		& 134 & 156 & 11 & 35   & 2\\
Seoul	  	& Se		& 420 & 466 & 12 & 35   & 3\\
Shanghai	  	& Sh		& 239 & 264 & 11 & 35   & 3\\
Mexico City	& MC	& 147 & 164 & 11 & 31   & 2\\
Berlin		& Ber	& 170 & 282 & 10 & 29   & 2\\
Chicago		& Ch		& 167 & 222 & 8   & 25   & 2\\
Osaka		& Os		& 108 & 123 & 9   & 24   & 2\\
Beijing		& Bei	& 163 & 176 & 13 & 21   & 4\\
Hong Kong	& HK		& 84   & 87   & 10 & 12   & 4\\
\bottomrule[1pt]
\end{tabular*}
\caption{{\bf Network characteristics of the largest connected
    component for the 15 largest metropolitan systems in the world.}
  The number of routes $N$ and connections $K_{\mathrm{tot}}$, respectively, yield nodes and edges in the
  dual space. We list cities from most connections to fewest
  connections between different lines. The number of connections $K_{\mathrm{tot}}$ is
  the key quantity from the perspective of information processing. (See
  the right panel of Fig.~2.) \emph{P-diameter} indicates the network diameter in dual space. It is equal to 2 for 10 of the 15 networks, and one additionally obtains a value of 2 in Paris if one cuts ``3bis'' (a four-stop line). 
}
\label{table1}
\end{table}

\begin{table}[h!]
\centering
\ra{1.3}
\begin{tabular*}{.8\textwidth}{@{\extracolsep{\fill}}lccccccc@{}}
\toprule[1pt]
Network & Nodes & Edges & $N$ & $K_{\mathrm{tot}}$ \\
\hline
NYC Metro		&412		&512		&20		&162 \\
NYC Bus			&5306	&8435	&309		&6092 \\
NYC multilayer		&5332	&8804	&330		&8461 \\
\hline
Paris MRT		&629		&765		&28		&162 \\
Paris Bus			&3842	&5567	&277		&2708 \\
Paris multilayer		&4037	&6142	&305		&4292 \\
\hline
Tokyo Metro		&217		&262		&13		&56 \\
Tokyo Bus			&1359	&1663	&153		&1275 \\
Tokyo multilayer	&1422	&1908	&166		&1831 \\
\bottomrule[1pt]
\end{tabular*}
\caption{{\bf Network characteristics of the Bus-Metro multilayer
    networks.} Similarly to Table~1, we show the number of routes
  $N$ and connections $K_{\mathrm{tot}}$, and the nodes and edges of
  the dual space. We note that there is a difference of an order of
  magnitude between the dimension of the Metro and the Bus layers,
  which represents a huge jump in complexity challenging people's ability to navigate the multilayer transport networks.  
}
\label{table4}
\end{table}

\clearpage

\begin{table}[h!]
\centering
\ra{1.3}
\begin{tabular*}{.5\textwidth}{@{\extracolsep{\fill}}lcrrr@{}}
\toprule[1pt]
	 & $C=1$ \ & $C=2$ \ & $C=3$ \ \\
\hline
NYC	 & 85.2\%	& 14.8\%	& 0\%	\\
Paris MRT & 43.0\%	& 48.5\%	& \ 8.5\%	\\
Tokyo& 72.8\% & 28.2\%	&0\%	\\
\bottomrule[1pt]
\end{tabular*}
\caption{{\bf Structure of Simplest Paths in Three Metro Systems.} We compare the number of connections in the simplest paths for the metro systems of the three megacities (New York City, Paris, and Tokyo) that we consider in detail. Only Paris has paths with more than 2 connections. A negligible fraction (not displayed) of paths have 4 or 5 connections.}
\label{table2}
\end{table}

\begin{table}[h!]
\centering
\ra{1.3}
\begin{tabular*}{.3\textwidth}{@{\extracolsep{\fill}}lrcc@{}}
\toprule[1pt]
	& C=1  &  C=2  \\
\hline
NYC	 & 99.4\% & 0.5\%	 \\
Paris & 86.4\%	& 13.6\%	 \\
Tokyo& 79.9\%	& 20.1\%	 \\
\bottomrule[1pt]
\end{tabular*}
\caption{{\bf Length proportions of the paths with $\bar S<8.1$ bits.} For the three megacities that we consider in this paper, about 20\% of the trips have an information entropy that is lower than the threshold of 8.1 bits. Such trips predominantly have only a single connection. When there are more, the starting route has a limited number of connections (see Fig.~\ref{tokyo20perc}).
}
\label{table3}
\end{table}

\clearpage

\begin{figure*}[h!]
\centerline{
\includegraphics[width=0.78\linewidth]{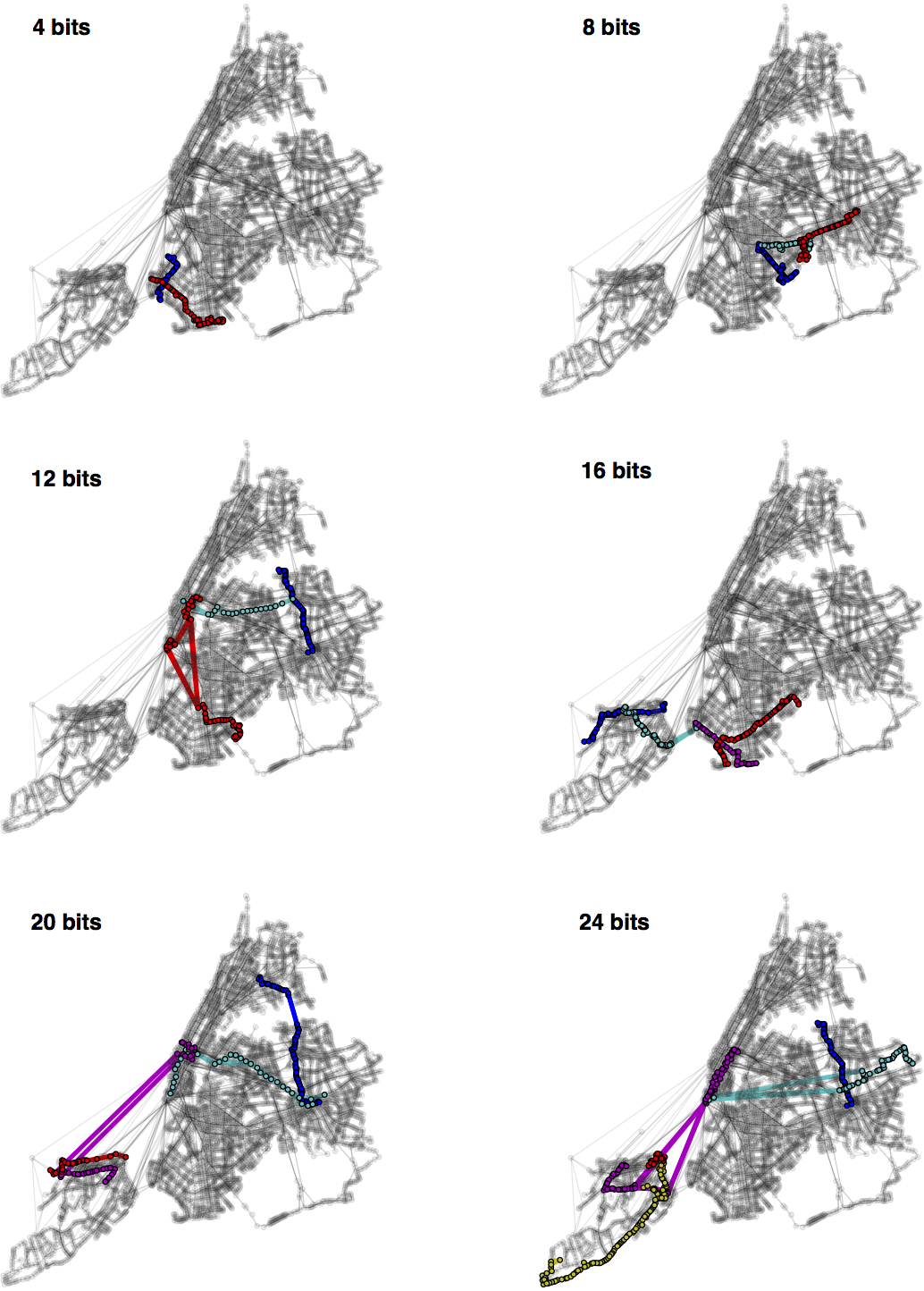} 
}
\caption{{\bf Examples of paths with growing $\bar S(s,t)$.} For the New York City multilayer network, we show examples with increasing complexity: $\bar S(s,t)$ ranges from 4 bits to 24 bits. We color the starting bus line $s$ in blue and the destination line $t$ in red.}
\label{examples}
\end{figure*}

\clearpage

\begin{figure*}[h!]
\centerline{ \includegraphics[width=0.9\linewidth]{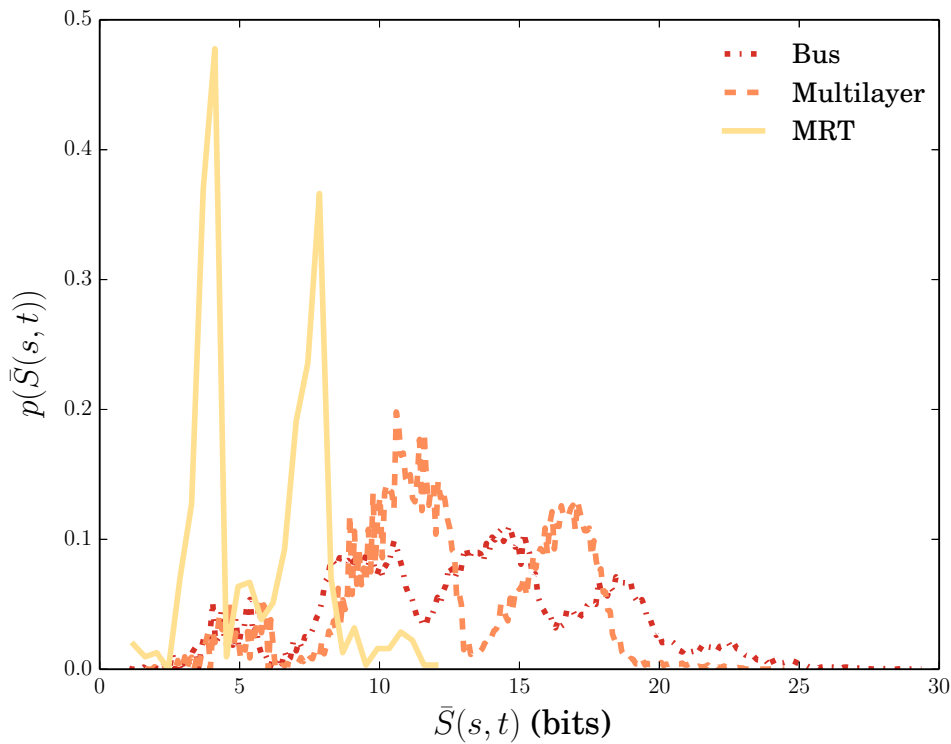} }
\caption{{\bf Entropy distribution for MRT layer, bus layer, and complete multilayer network in Paris.} For the multilayer network, we restrict the distribution to trips whose origin and destination are each in the bus layer.
We see that the effect of multiplexity on the bus layer is to shift the peaks to the right, and we also obtain larger peaks for smaller values of $C$ (see Fig.~\ref{multilayerEffect}).
}
\label{multilayerDistributions}
\end{figure*}

\begin{figure*}[h!]
\centerline{ \includegraphics[width=0.45\linewidth]{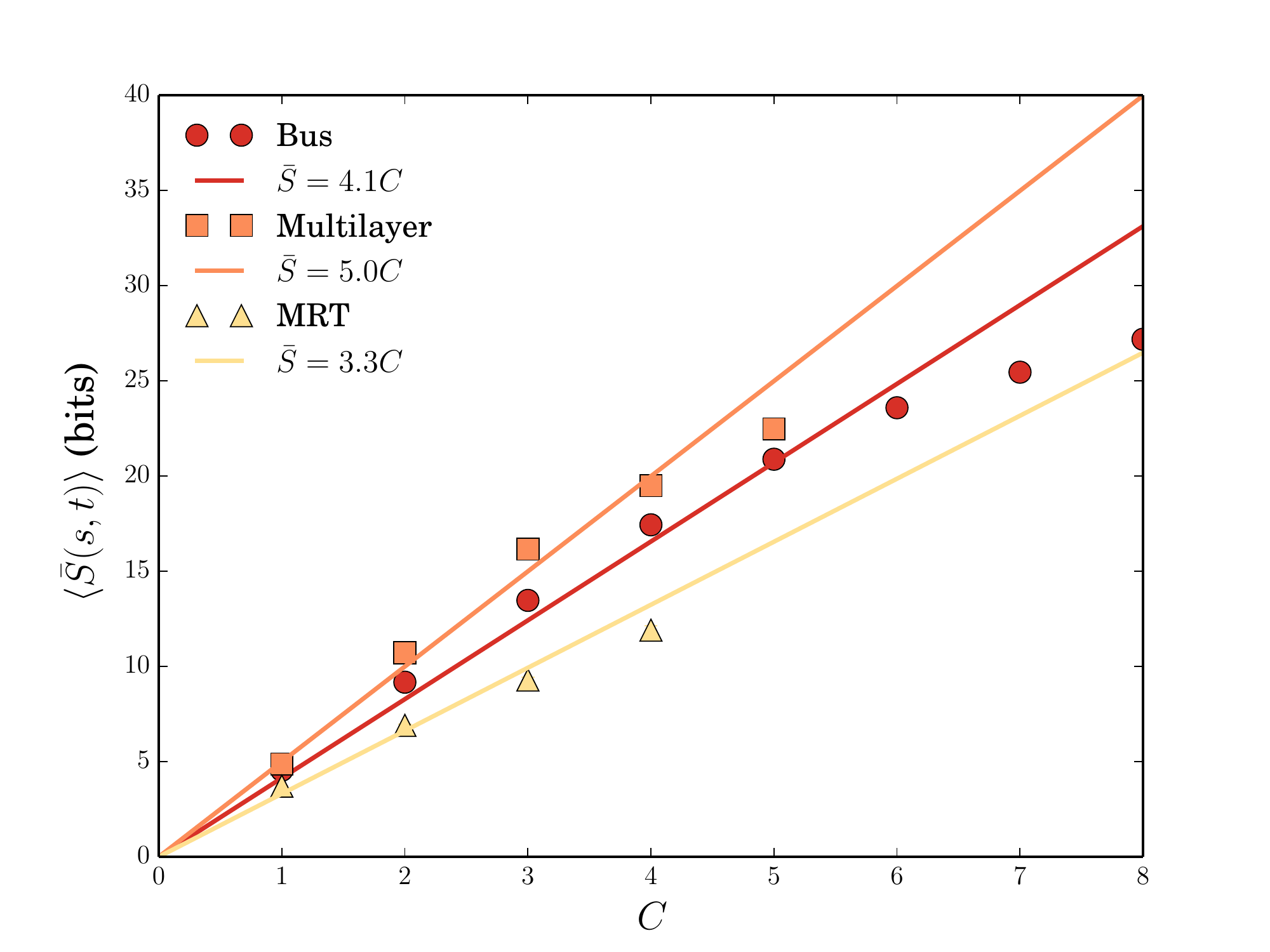}
\qquad \includegraphics[width=0.45\linewidth]{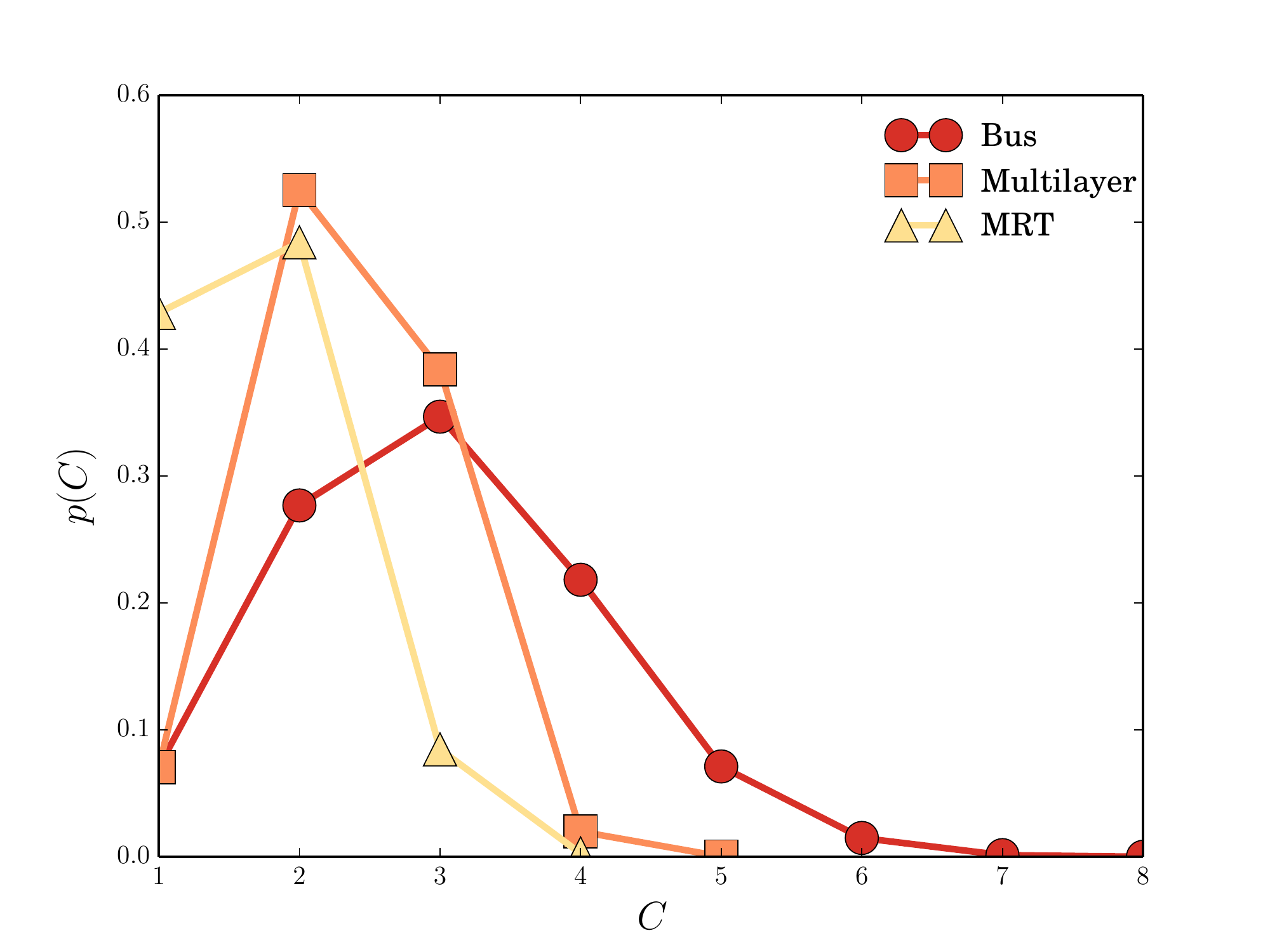}
 }
\caption{{\bf Effects of multiplexity.} {\bf (Left)} The values of $\bar S$ grow with $C$.
The growth is larger for the multilayer network (in the sense that it
is characterized by the highest value of $\langle k \rangle$), smaller
for the bus layer (where the lines have fewer connections), and is
smallest for the MRT layer. {\bf (Right)} Conversely, the mean path length is smaller for the bus monolayer network than for the multilayer network, in which the bus service interacts with the (longer-range) lines in the MRT layer.}
\label{multilayerEffect}
\end{figure*}

\begin{figure*}[h!]
\centerline{ \includegraphics[width=0.8\linewidth]{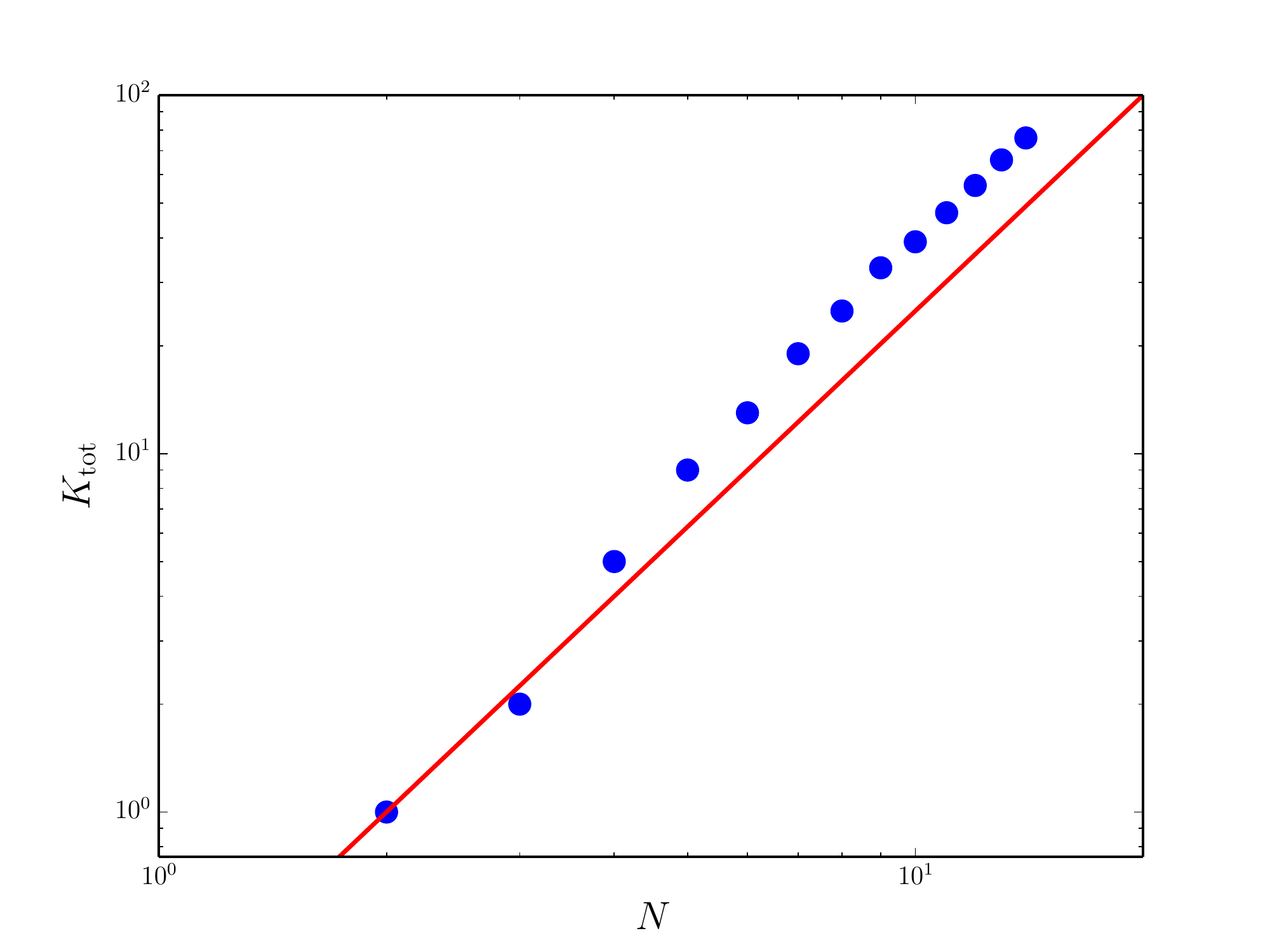} }
\caption{{\bf Growth of the Paris Metro Network.} Letting the network grow with its historical progression from line 1 to line 14, we see that the number of connections in the dual space $K_\mathrm{tot}$ (blue dots) grows similarly to a lattice (red line), which would have $(N/2)^2$ intersections.
}
\label{parisGrowth}
\end{figure*}

\begin{figure*}[h!]
\centerline{
\includegraphics[width=0.8\linewidth]{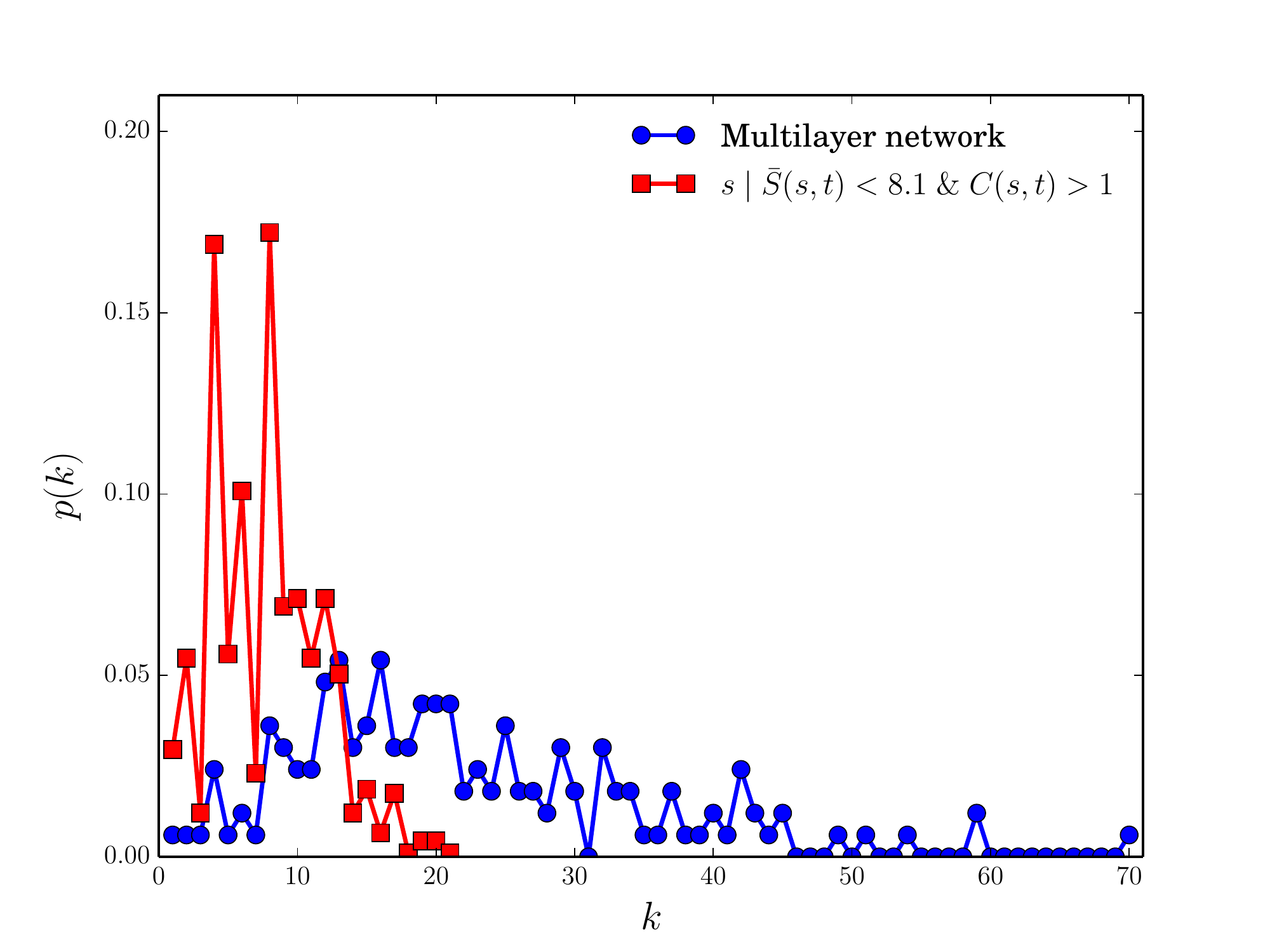} 
}
\caption{{\bf Dual-space degree of low-information starting points for
    paths with $C=2$ in the Tokyo multilayer network.}  As one can see in
  Table~\ref{table3}, most of the trips below the
  cognitive limit have $C=1$ connections. The trips below the
  cognitive threshold with $C>1$ connections are characterized by a
  low connectivity of the origin route. We show this feature for the
  Tokyo case. We see that $C=2$ for $20.1\%$ of the trips that are below the threshold of
  $8.1$ bits. For this fraction of trips $s\to t$, the
  degrees $k$ in the dual space of origin routes $s$ (red
  squares) are small in comparison to the degrees of all routes on the
  whole multilayer network (blue circles).}
\label{tokyo20perc}
\end{figure*}

\begin{figure*}[h!]
\centerline{ \includegraphics[width=0.45\linewidth]{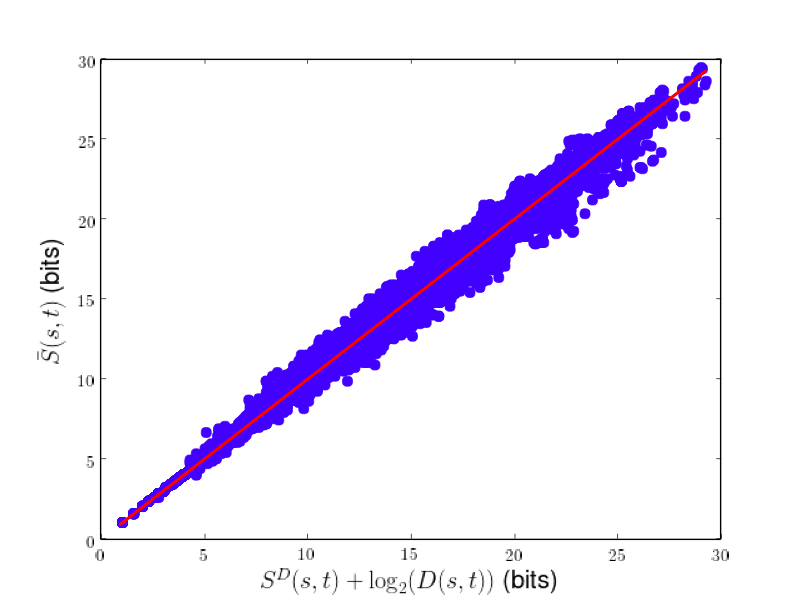}
\qquad \includegraphics[width=0.45\linewidth]{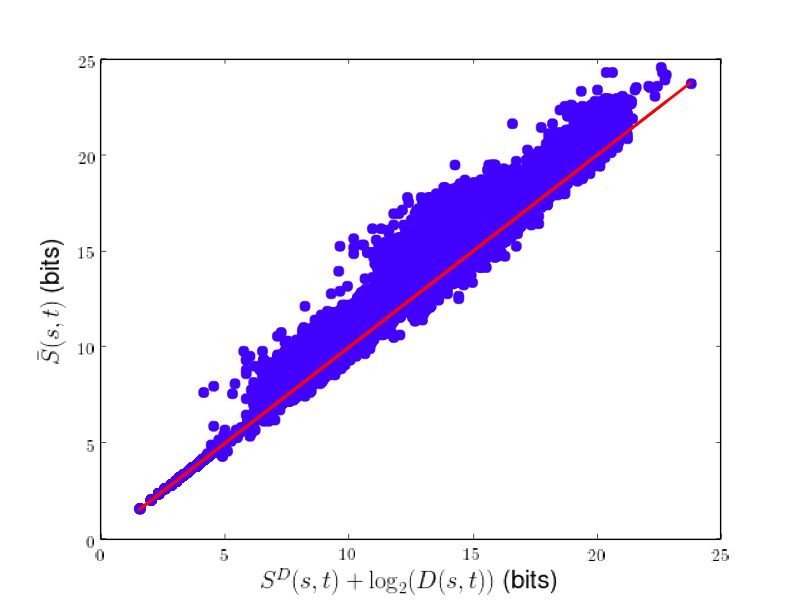}
 }
\caption{{\bf Empirical validation of Eq.~(7).} Comparing the (left) bus monolayer network to the (right) multilayer network in Paris, we note that including the metro-rail-tram (MRT) layer yields larger fluctuations.
The mean square deviation is 0.34 bits for the bus layer and 0.82 bits of the multilayer network that contains both bus and MRT modes. This suggests that for the same route pair $(s,t)$, different paths become optimal for different origins $i$ and destinations $j$. 
}
\label{fluctuations}
\end{figure*}

\begin{figure*}[h!]
\centerline{
\includegraphics[width=0.8\linewidth]{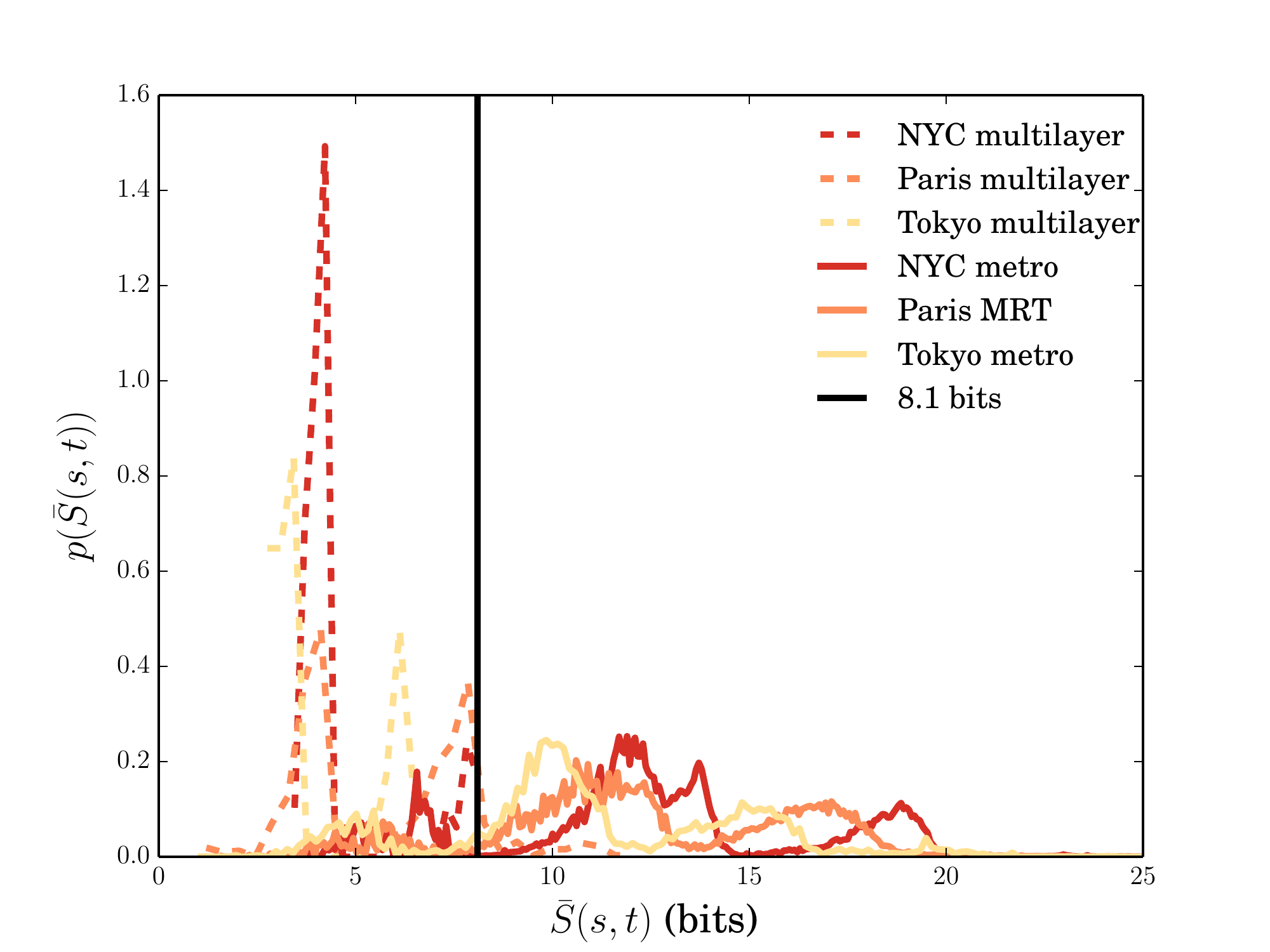} 
}
\caption{{\bf Information entropy of multilayer networks}. This figure represents the probability density distributions of $\bar S(s,t)$. In Fig.~4 of the main text, we show the associated cumulative distributions. Similar to Fig.~4, we associate one layer to bus routes and another to metro lines. The solid curves are associated with multilayer networks that include a metro layer for New York City, Paris, and Tokyo. The dashed curves are associated to all possible paths in a metro layer. We observe that every distribution is characterized by a peak structure, and every peak is associated to a number $C$ of connections.}
\label{}
\end{figure*}

\end{document}